\documentclass[aps,twocolumn,showpacs,nofootinbib,superscriptaddress]{revtex4}
\usepackage{graphicx}
\usepackage{amsmath}
\usepackage{amsfonts}
\usepackage{amssymb}
\usepackage{hyperref}

\newcommand{\ket}[1]{|{#1}\rangle}
\newcommand{\bra}[1]{\langle{#1}|}

\newcommand{\beq}{\begin{equation}}
\newcommand{\eeq}{\end{equation}}
\newcommand{\beqa}{\begin{eqnarray}}
\newcommand{\eeqa}{\end{eqnarray}}

\begin{document}

\title{Decoherence induced by a dynamic spin environment (I): The universal regime}

\author{Cecilia Cormick}
\affiliation{Departamento de F\'\i sica, FCEyN, UBA,
Ciudad Universitaria Pabell\'on 1, 1428 Buenos Aires, Argentina}

\author{Juan Pablo Paz}
\affiliation{Departamento de F\'\i sica, FCEyN, UBA,
Ciudad Universitaria Pabell\'on 1, 1428 Buenos Aires, Argentina}

\date{\today}

\begin{abstract}
This article analyzes the decoherence induced on a single qubit by the 
interaction with a spin chain with nontrivial internal dynamics (XY-type 
interactions). The aim of the paper is to study the existence and properties
of the so-called universal regime, in which the decoherence time scale becomes independent 
of the strength of the coupling with the environment. It is shown that although such regime 
does exist, as previously established by Cucchietti \textit{et al} in {\em Phys. Rev. A}, 75:032337 (2007), 
it is not a clear signature of a quantum phase transition in the environment. In fact, this kind of universality also exists in the absence of quantum
phase transitions. A universal regime can be related to the existence of an energy scale separation between the Hamiltonian of the environment and the one 
characterizing the system-environment interaction. The results presented also indicate that in the strong coupling regime the quantum phase transition does not produce 
an enhancement of decoherence (as opposed to what happens in the weak coupling regime). 
\end{abstract}

\date{\today}
\pacs{03.65.Yz, 03.67.Lx, 05.70.Fh} 
\maketitle

\section{Introduction}

Decoherence \cite{Zeh-1973, PazZ00, Zurek03} is the main obstacle that prevents us from taming the quantum world taking advantage of its remarkable properties. In fact, uncontrolled interactions of a quantum system with its environment result in a dramatic suppression of quantum phenomena such as interference and entanglement within the system. The understanding of this dynamical process is essential in order to devise quantum information processors \cite{NielsenC00}, not only to find ways to minimize (or control) the effects induced by the environment but also to design appropriate error correction techniques \cite{Shor95, Steane96, Gottesman00} or error protection strategies (using, for example, decoherence-free subspaces to encode quantum information \cite{lidar-1998-81}). 

In recent years the study of spin-bath environments has attracted much attention [10--19] 
since for some qubit systems such type of environment could provide a quite realistic model of the relevant decoherence process (when the model of an environment as a collection of non-interacting harmonic oscillators fails to describe the observed behaviour). The study of decoherence induced by an environment with nontrivial dynamics has attracted special interest. For bosonic environments, various recent studies focus on the influence of non-linear and chaotic effects. In some cases, such effects seem to enhance the capability of a given reservoir to efficiently induce decoherence \cite{ErmannPS-2006, Bluhme-KohoutZ-2003}. 
For spin baths, the problems considered so far include, for example, the effect of intra-environment interactions for a low-temperature spin bath \cite{tessieri-2003-36} and the decay of coherence caused by an environment of independent spins, in regimes dominated by the interaction Hamiltonian or the Hamiltonian of the system \cite{cucchietti-2005-72}. Other authors have studied the consequences of quantum phase transitions in the environment, in the central spin model where a qubit interacts homogeneously with all spins in a chain with XY Hamiltonian \cite{quan-2006-96, cucchietti-2006} and in a more general case where the qubit interacts with an arbitrary number of sites in the chain \cite{rossini-2007-75}. The loss of entanglement of a system formed by two qubits has been examined in cases with homogeneous couplings to the same XY chain \cite{jing-2006}, and with nonhomogeneous couplings to chaotic, integrable, and mixed environments \cite{Pineda:012305}. Other effects were analyzed using numerical simulations (that are limited by the exponential scaling of the required resources) \cite{YuanKD-2007, DeRaedtD-2004}.  

As remarked by Cucchietti, Fernandez-Vidal and one of us in \cite{cucchietti-2006}, another interesting aspect of decoherence studies is the following: it could be possible to profit from the decoherence process using the quantum system as a probe to learn about properties of the environment. In  \cite{cucchietti-2006} an example was proposed and analyzed: the main idea was to use one qubit as a detector for a quantum phase transition taking place in the environment, as this would become manifest in certain universality features of the decoherence process \cite{cucchietti-2006}. In this paper we shall examine more carefully the regime of universal decoherence that was the focus of that proposal. The system we will consider is formed by one qubit interacting with an environment given by a chain of spins with XY Hamiltonian. We will study the existence and features of the universal regime in this problem and show that the conclusions in \cite{cucchietti-2006} are not generic and must be taken with a grain of salt. We will show that the universal regime is not always Gaussian, as had already been noted in \cite{rossini-2007-75}. After discussing the properties of the Gaussian and non-Gaussian universal regimes we will analyze the connection between the universality and the existence of a quantum phase transition in the environment. 

The paper is organized as follows: in Section \ref{sec:themodel} we introduce the model. We describe both the system and the environment by defining their Hamiltonians, and we present the main formulas we will use to determine the decay of quantum coherence. Section \ref{sec:gaussianregime} is devoted to the regime of universal decoherence in the Gaussian case considered in \cite{cucchietti-2006}; in Section \ref{sec:nongaussianregime} we will study the non-Gaussian case; in Section \ref{sec:universality} we will discuss the reasons for the existence of the universal regime. Finally, in Section \ref{sec:conclusions} we summarize our results.

\section{The model: one central qubit interacting with a spin chain}
\label{sec:themodel}

\begin{figure}[!hbt]
\begin{center}
\includegraphics[width=0.2\textwidth]{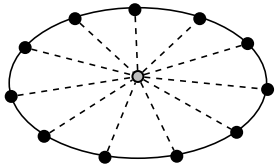}
\caption{The model: a one-qubit system equally coupled to all sites of a spin chain.}
\label{fig:1spin}
\end{center}
\end{figure}

We will study the decoherence induced on a spin $1/2$ particle
(which we shall call ``the system'' or ``the qubit'') by the coupling to an environment formed by a chain of $N$ spin $1/2$ particles. We neglect the self-Hamiltonian of the system, and consider that the qubit interacts equally with all the spins in the chain (Figure \ref{fig:1spin}). The Hamiltonian of the environment chain will be taken as:
\beq
H_C=-\sum_j\left\{ \frac{1+\gamma}{2} X_j X_{j+1} + \frac{1-\gamma}{2} Y_j Y_{j+1} + \lambda Z_j\right\}
\eeq
where periodic boundary conditions are imposed, and the three Pauli operators acting on the $j$-th site of the chain are denoted as $X_j,Y_j,Z_j$. The parameter $\gamma$ determines the anisotropy in the $x-y$ plane and $\lambda$ gives a magnetic field in $z$ direction ($\gamma=1$ corresponds to the Ising chain with transverse field). 
This model is critical for \mbox{$\gamma=0$}, $|\lambda|<1$ and for $\lambda=\pm1$, which corresponds (in the limit $N\to\infty$) to a quantum phase transition from a ferromagnetic to a paramagnetic phase. Throughout the paper, we shall discuss the effects of this phase transition on the decoherence of the system. 

The interaction Hamiltonian is chosen as:
\beq
H_{int}=-g \ket{1}\bra{1} \otimes \sum_j Z_j.
\eeq 
with $\ket{0}$ and $\ket{1}$ the two eigenstates of the Pauli operator for 
the qubit $Z_S$. Thus, depending on the state of the system, the environment evolves with a different effective Hamiltonian $H_{a},~a=0,1$, given by:
\beq
H_a=H_C-a g \sum_j Z_j.\nonumber
\eeq  
which is an effective change of the external field as $\lambda\to\lambda^{(a)}=\lambda+ag$.

We assume the initial state of the universe formed by the system and the environment to be pure and separable (i.e., the environment is not correlated with the system):
\beq
\rho_{SE}(0)= \ket{\psi}\bra{\psi} \otimes \ket{E_0}\bra{E_0}
\eeq
with $\ket{\psi}=\alpha\ket{0}+\beta\ket{1}$. Our goal is to study the evolution of the reduced density matrix of the system (obtained from the state of the universe by tracing out the environment). Because of the special form of the Hamiltonian, the temporal dependence of the reduced density matrix $\rho$ can be formally obtained as follows. In the computational basis of the system (formed by the eigenstates of $Z_S$), $\rho$ can be written as:
\beq
\rho(t)=\sum_{ab=0,1}\rho_{a,b}(t) \ket{a}\bra{b}.
\eeq
The evolution of the matrix elements of $\rho$ is given by:
\beq
\rho_{a,b}(t)=\rho_{a,b}(0) \bra{E_0}e^{iH_{b}t} e^{-iH_{a}t}
\ket{E_0}
\eeq
(we are using units such that $\hbar=1$). 
As the total Hamiltonian commutes with $Z_S$, the diagonal terms $\rho_{a,a}$ remain constant. The amplitude of each off-diagonal term is instead multiplied by the overlap between two states of the environment that correspond to the two different evolutions of the chain according to the different system states. To analyze the decoherence induced by the spin chain we will consider the square modulus of this factor, which following \cite{rossini-2007-75} we shall call the Loschmidt echo: 
\beq
L(t) = |\bra{E_0}e^{iH_0 t} e^{-iH_1 t}\ket{E_0}|^2.
\eeq
This echo is simplified if we assume the environment to be initially in its ground state. In such a case, one of the evolution operators in the expression acts trivially, and the echo is then equal to the survival probability of the initial state after being evolved with the effective Hamiltonian $H_1$.

While the initial state is pure and separable, as a consequence of the interaction the qubit becomes entangled with the chain, and its reduced density matrix becomes mixed. The purity of the qubit as a function of time can be computed from the Loschmidt echo as:
\beq
Tr(\rho^2(t))=1-2|\alpha\beta|^2(1-L).
\eeq

To find the solution $L(t)$ for this problem we first note that the Hamiltonians $H_{a}$ of the chain can be mapped onto a fermion system by means of the Jordan-Wigner transformation \cite{LiebSM-1961}:
\beqa
X_j&=&exp\left\{i\pi\sum_{k=1}^{j-1}c_k^\dagger c_k\right\} (c_j+c_j^\dagger)\\
Y_j&=&i~exp\left\{i\pi\sum_{k=1}^{j-1}c_k^\dagger c_k\right\} (c_j-c_j^\dagger)\\
Z_j&=&2c_j^\dagger c_j-1. \label{eq:transfZ}
\eeqa
Using this, up to a correction term associated to boundary effects, the Hamiltonians can be written as:
\beqa
H_a&=&-\sum_j (c_j^\dagger c_{j+1} +c_{j+1}^\dagger c_j) + \gamma (c_j^\dagger c_{j+1}^\dagger +c_{j+1} c_j) +\nonumber\\
&&+ \lambda^{(a)} (2c_j^\dagger c_j-1). 
\eeqa

Because the qubit interacts homogeneously with the chain, the Hamiltonians $H_a$ are translationally invariant and can be diagonalized by a standard method. First we  
Fourier transform the creation and annihilation operators $c, c^\dagger$. Then
we define new operators by a Bogoliubov transformation that preserves momentum, mixing only the Fourier-transformed operators $\tilde c_k$ and $\tilde c^\dagger_{-k}$: 
\beq 
\tilde c_k = \eta_k^{(a)} \cos\left(\frac{\varphi^{(a)}_k}{2}\right) - \eta_{-k}^{\dagger(a)} \sin\left(\frac{\varphi^{(a)}_k}{2}\right).\label{Bogoliubov}
\eeq
The Bogoliubov coefficients obey the relation: 
\beq \label{eq:theangles}
\tan(\varphi^{(a)}_k) = \frac{\gamma ~\sin(2\pi k/N)}{\lambda^{(a)}+\cos(2\pi k/N)}
\eeq
and the particle energies are:
\beq 
E^{(a)}_k = 2\left( (\gamma ~\sin(2\pi k/N))^2 + (\lambda^{(a)}+\cos(2\pi k/N))^2 \right)^{1/2}.
\eeq

The mixing between creation and annihilation operators thus depends on angles that change when the external field is varied as a consequence of the interaction with the central system. Because this mixing preserves momentum, the square modulus of the factor modulating the off-diagonal terms of $\rho$ can be factorized, giving: 
\beq \label{eq:mainformula}
L (t) = \prod_{1\leq k<N/2} \left(1-\sin^2(\delta \varphi_k) \sin^2(E^{(1)}_k t) \right)
\eeq
with $\delta \varphi_k=\varphi^{(1)}_k-\varphi^{(0)}_k$.

\section{The Gaussian universal regime}
\label{sec:gaussianregime}

\begin{figure}[!hbt]
\begin{center}
\includegraphics[width=0.45\textwidth]{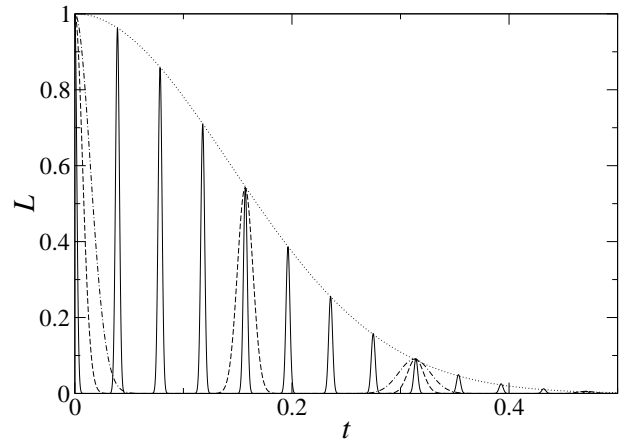}
\caption{The echo as a function of time for the case considered by Cucchietti \textit{et al}: a central one-qubit system interacting equally with all sites of a spin chain for $\gamma=1$, $\lambda=0$ (we take $N=100$). In this case, the universal regime is already reached for the curves shown, corresponding to $g=5$ (dash-dotted), 10 (dashed), 40 (full); the dotted line indicates the universal (Gaussian) envelope.}
\label{fig:pcv}
\end{center}
\end{figure}

The problem under study was analyzed in \cite{cucchietti-2006} for an Ising ($\gamma=1$) environment chain. In that paper, the existence of a universal decoherence regime was discovered, for $\lambda<1$ and \mbox{$g>1$}. This regime was later observed and discussed by other authors \cite{rossini-2007-75}. The universal regime found in \cite{cucchietti-2006} (illustrated in Figure \ref{fig:pcv}) is characterized by the fact that the echo has a Gaussian envelope whose width is independent of the strength of the coupling to the environment, parametrized by $g$. The existence of a universal (i.e., $g$-independent) Gaussian envelope was shown in \cite{cucchietti-2006} to be a consequence of certain properties of the distribution of the Bogoliubov coefficients appearing in equation (\ref{Bogoliubov}). In particular, it was related to the fact that for $\gamma=1$, \mbox{$\lambda\simeq0$}, and large enough $g$, the angles $\delta \varphi_k$ are uniformly distributed. 

\begin{figure}[!hbt]
\begin{center}
\includegraphics[width=0.45\textwidth]{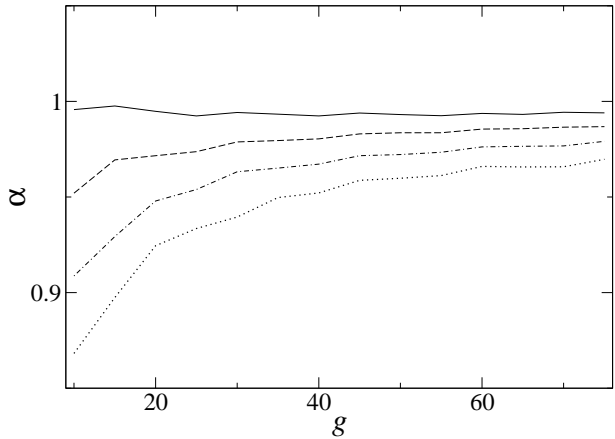}
\caption{The echo for a central one-qubit system interacting equally with all sites of a spin chain has, for the case \mbox{$\gamma=1$}, \mbox{$\lambda<1$}, $g>1$, a Gaussian envelope of the form $\exp(-\alpha t^2 N/4)$. The value of $\alpha$ is plotted as a function of $g$ for  $\lambda=0$ (full), 0.3 (dashed), 0.6 (dash-dotted) and 0.9 (dotted), and a chain length $N=100$. The fits are taken considering peaks for which the echo is over $1/e$.}
\label{fig:alpha}
\end{center}
\end{figure}

In what follows we shall examine in more detail the features of this Gaussian universal regime. The observed universality is reached when the strength $g$ of the interaction is above a certain threshold. The value of this threshold seems to be independent of the length of the chain $N$, but it depends on $\lambda$. The Gaussian envelope behaves as $\exp(-\alpha t^2 N/4)$; the asymptotic value of $\alpha$, the parameter determining the decay width, is roughly 1. The dependence of $\alpha$ on the strength of the perturbation $g$ is analyzed in Figure \ref{fig:alpha}, for $\lambda$ between 0 and 0.9. We must note, however, that as $\lambda$ approaches 1 the envelope acquires a slowly decaying tail, so that the Gaussian fit is not good at long times. It is clear from the Figure that for $\lambda=0$ (top curve), the universality is reached faster than for $\lambda=0.9$ (bottom curve). The dependence of the threshold on $\lambda$ is not surprising: by considering $\lambda$ close to unity we are taking an initial state which approaches the critical vacuum state from below. In such a case, the effect of the perturbation must be smaller than for $\lambda\simeq0$, as the overlap between the two ground states of the effective Hamiltonians is expected to be larger. Nevertheless, it must be noted that the dependence of $\alpha$ on $g$ is rather weak. As seen in the Figure, for $\lambda=0.9$ the changes in $\alpha$ are only of the order of $10\%$  for $g$ varying from $10$ to $75$. Besides, the fact that $\alpha$ decreases with $\lambda$ indicates that, contrary to the weak coupling case, the proximity to the quantum phase transition is not responsible for an enhancement of decoherence. Decoherence is actually weaker for larger $\lambda$, because the spins tend to align with the external field. 

The values of $\alpha$ shown in Figure \ref{fig:alpha} correspond to Gaussian fits of the peaks in the evolution of the echo as a function of time. They can also be approximated by a simple analytical formula as follows. In the expression of the echo (\ref{eq:mainformula}) we consider all energies to be $E^{(1)}_k = E+\Delta_k$ with $\Delta_k \ll E$ (which in this model is satisfied for large $g$). All the factors in the echo thus oscillate with almost the same frequency, the differences $\Delta_k$ being responsible for the echo decay. Evaluating near the peaks, $t=n\pi/E+\delta t$, and using Taylor expansions in $\delta t$ and $\Delta_k$, we find that the frequency of the peaks corresponds to an energy $E$ given by:
\beq \label{eq:E_alfa_analitico}
E = \frac{\sum_k \sin^2(\delta \varphi_k) E^{(1)}_k}{\sum_k \sin^2(\delta \varphi_k)}
\eeq
and the value of the echo at these peaks can be approximated by $\exp(-\alpha t^2 N/4)$ with:
\beq \label{eq:alfa_analitico}
\alpha \frac{N}{4}= \sum_{1\leq k<N/2} \sin^2(\delta \varphi_k) (E^{(1)}_k -E)^2.
\eeq

\begin{figure}[!hbt]
\begin{center}
\includegraphics[width=0.45\textwidth]{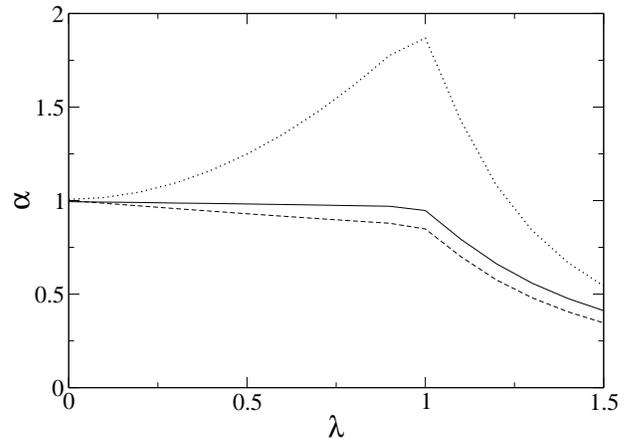}
\caption{The echo for a central one-qubit system interacting equally with all sites of a spin chain has, for the case \mbox{$\gamma=1$}, \mbox{$\lambda<1$}, $g>1$, a Gaussian envelope of the form $\exp(-\alpha t^2 N/4)$. The value of $\alpha$ is plotted as a function of $\lambda$ for $g=75$ (full), from Gaussian fits considering peaks for which the echo is over $1/e$. These are compared with our analytic approximation (dashed) and the one from the formula given in \cite{cucchietti-2006}.}
\label{fig:alpha_approx}
\end{center}
\end{figure}

This formula is similar to the one given in \cite{cucchietti-2006}. However, there is a substantial difference: the approach described in 
\cite{cucchietti-2006} basically takes into account the dispersion of the energies about the mean value. This gives a good approximation for the case $\lambda=0$, which was the one analyzed in that paper. But it fails to reproduce the behaviour of the echo as $\lambda$ is increased close to its critical value, wrongly predicting an enhancement of decoherence by the quantum phase transition in the strong coupling case (as was asserted in \cite{YuanZL-2007}). This effect, which can indeed be found for weak coupling, is not observed here for large $g$. The weight $\sin^2(\delta\phi_k)$ inside the sum appearing in (\ref{eq:E_alfa_analitico}) is essential in the estimation of the decay width of the Gaussian for $\lambda\simeq1$. In Figure \ref{fig:alpha_approx} we show the comparison between our approximation, the one given in \cite{cucchietti-2006}, and the Gaussian fit of the results. Both approximations are good for $\lambda\simeq0$. For larger values of $\lambda$ our approximation underestimates decoherence, but displays the right behaviour at the critical point. 

\section{A non-Gaussian universal regime}
\label{sec:nongaussianregime}

\begin{figure}[!hbt]
\begin{center}
\includegraphics[width=0.45\textwidth]{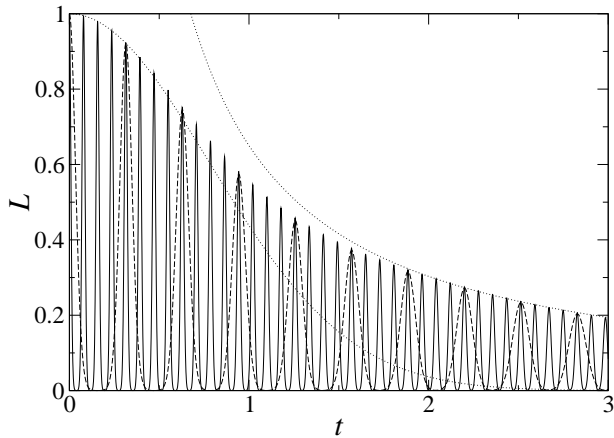}
\caption{The echo as a function of time for a central one-qubit system interacting equally with all sites of a spin chain of length $N=100$, for $\gamma=0.1$, $\lambda=0$. In this case, the universal regime is also reached for the curves shown, corresponding to $g=5$ (dashed), 20 (full), but the envelope is no longer Gaussian. The dotted lines indicate the Gaussian behaviour at short times, followed by a power-law decay ($\sim t^{-{1.1}}$).}
\label{fig:env_gamma01}
\end{center}
\end{figure}

It is interesting to notice that the Gaussian nature of the envelope disappears if we consider an environment with a more complex evolution. In fact, the Gaussian envelope is  seen in the Ising ($\gamma=1$) case. For other values of the anisotropy parameter $\gamma$ there is a Gaussian regime for short times which is followed by power-law decay (Figure \ref{fig:env_gamma01}). It is worth mentioning that the transition from a Gaussian to a power-law decay was also observed for the decoherence induced by other spin baths when changing the Hamiltonian of the environment \cite{zurek-2007-38, cucchietti-2005-72}. 

Nevertheless, the universal behaviour is similar to the Gaussian case, namely, once the interaction strength $g$ is over a certain threshold the envelope becomes $g$-independent. Even though the study of the envelope is not as simple as for $\gamma=1$, as it cannot be characterized by a single parameter, it is possible to see that the value of the threshold increases as $\lambda$ approaches unity, while it is not specially sensitive to $\gamma$ or $N$, in accord with the previous results.

\begin{figure}[!hbt]
\begin{center}
\includegraphics[width=0.45\textwidth]{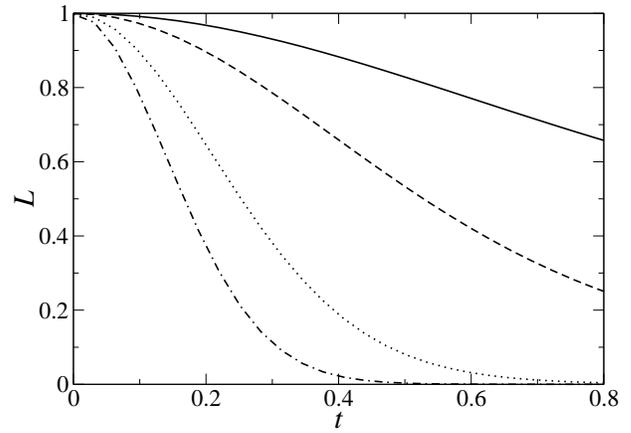}
\caption{The envelope of the echo for a central qubit interacting with all sites of a spin chain of length $N=100$ ($\lambda=0$, $g=50$). Different curves correspond to $\gamma=~$0.1 (full), 0.2 (dashed), 0.5 (dotted), 1 (dash-dotted). The Gaussian regime is only seen for $\gamma=1$, and the width of the envelope decreases with $\gamma$.}
\label{fig:env_distintosgamma}
\end{center}
\end{figure}

Taking into account the analysis in \cite{cucchietti-2006}, the non-Gaussian character of the envelope is not surprising, as for $\gamma\simeq0$ the conditions that were shown to lead to the Gaussian shape are not fulfilled: namely, the angles $\delta\varphi_k$ in (\ref{eq:mainformula}) are not randomly distributed, as most of the Bogoliubov coefficients do not vary significantly when the transverse field changes. Besides, the width of the envelope increases as $\gamma\rightarrow 0$, as seen in Figure \ref{fig:env_distintosgamma}. This should not be confused with the fact that the echo is exactly equal to $1$ for $\gamma=0$, as in this case the perturbation commutes with the chain Hamiltonian ($\gamma\to0$ is a singular limit). For small but nonzero $\gamma$, it is the top of the envelope that approaches 1 as $\gamma$ is decreased. This can be qualitatively explained by looking at the angles in (\ref{eq:theangles}): when $\gamma$ is small, all angles are close to zero, except those for which the denominator is close to zero too (the low-energy excitations in $H_0$). Only these low-energy states have large angle variations when introducing the perturbation, and they are the only ones contributing to the echo decay in (\ref{eq:mainformula}). If we take a large $g$ and a small $\gamma$, all the relevant frequencies will be very similar (approximately $2g$), so that they will all oscillate at the same time, preventing the decay of the envelope. We thus find that in the strong perturbation regime, decoherence is weaker for small $\gamma$. 

It is worth noticing that this is opposite to the case of small perturbations, where the proximity to the critical region $\gamma=0, |\lambda|<1$ is responsible for the enhancement of decoherence \cite{quan-2006-96}; the argument we have just outlined is no longer valid here as for small $\gamma$ \textit{and} $g$, the frequencies involved will be small and with large dispersion. Once more, we find that the relation between decoherence and quantum phase transitions in the strong coupling regime is very different from the weak coupling case.

\section{Is the universal regime related to the quantum phase transition?}
\label{sec:universality}

In \cite{cucchietti-2006} it was argued that the universality of the envelope may be taken as an indicator of the existence of a quantum phase transition. However, the derivation of the universal Gaussian decay in \cite{cucchietti-2006} does not explicitly use the existence of such a transition but is based upon a simple hypothesis on the distribution of the eigenstates of the two effective Hamiltonians $H_0$ and $H_1$. From the evidence presented in that paper (confirmed by our studies here and by other authors \cite{rossini-2007-75}) the universal behaviour is observable for $g>1$, which, if $\lambda<1$, is enough to drive the system along the phase transition. 

\begin{figure}[!hbt]
\begin{center}
\includegraphics[width=0.45\textwidth]{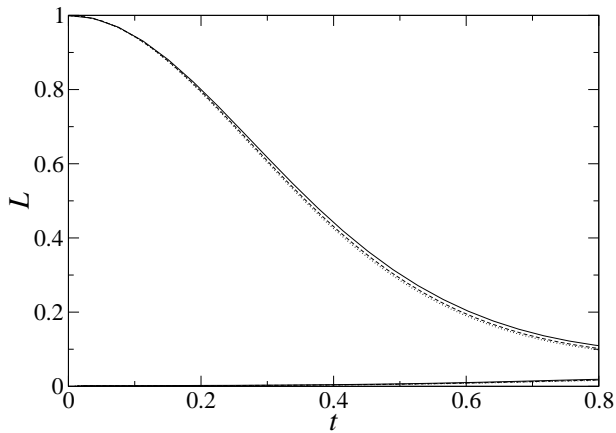}
\caption{The envelope of the echo as a function of time for a central one-qubit system,  with $N=100$, $\gamma=1$, $\lambda=2$. The envelopes for $g=$40 (full), 60 (dashed), 80 (dotted) are very alike, even though no phase transition is involved. We note that there is a ``bottom'' of the envelope which is not zero, and which gets closer to 1 when $\lambda$ is increased, as decoherence becomes weaker.}
\label{fig:env_lambda2}
\end{center}
\end{figure}

But from our studies a result becomes also evident: the existence of a universal regime is not a good indicator for detecting a quantum phase transition or structural unstability of the environment. This can be seen from the following example: by taking $\lambda>1$, a regime where the envelope becomes independent of the strength of the perturbation also appears (see, for example, Figure \ref{fig:env_lambda2}). The main difference between this regime and the one previously obtained for $\lambda<1$ is that in this new case decoherence is weaker (the ``bottom'' of the envelope is not $L=0$, and the decay is slower). In fact, decoherence becomes weaker and weaker as $\lambda$ is increased, which is entirely natural as the spins tend to align in the direction of the external field. But this new universal regime cannot be related to the existence of a phase transition, as both effective Hamiltonians in this case lie on the same side.

\begin{figure}[!hbt]
\begin{center}
\includegraphics[width=0.45\textwidth]{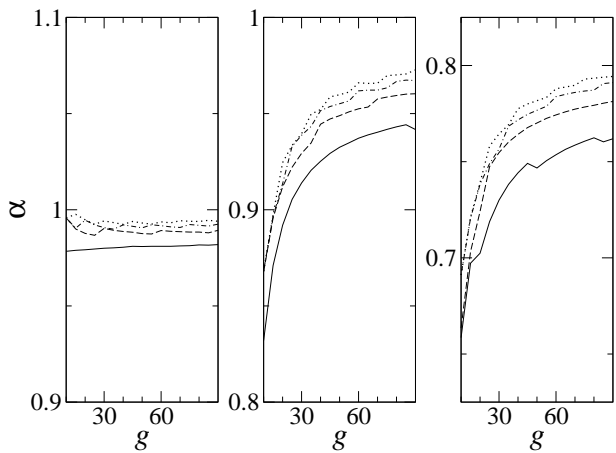}
\caption{The envelope of the echo for \mbox{$\gamma=1$} is adjusted at short times by a Gaussian fit $\exp(-\alpha t^2 N/4)$. The value of $\alpha$ is plotted as a function of $g$ for different values of $\lambda$: 0 (left), 0.9 (middle), and 1.1. In each plot the curves correspond to $N=40$ (full), 60 (dashed), 80 (dash-dotted) and 100 (dotted).}
\label{fig:universality}
\end{center}
\end{figure}

The fact that there are regimes of universal decoherence which are not related to quantum phase transitions was already discussed in \cite{cucchietti-2006} for complex systems. It is then not so surprising that in the case of the Ising chain the universal regime can be found for values of $\lambda$ such that no phase transition is involved. One might think that the transition is related not to the mere existence of the regime, but to scaling features, for instance how rapidly the regime is reached and how the threshold depends on $N$. In Figure \ref{fig:universality} we study this problem. We take $\gamma=1$ fixed and different values for $N$ and $\lambda$ (0, 0.9, and 1.1). We study the envelopes until they decay to a value of $1/e$; in this range and for these values of $\lambda$ the Gaussian shape, $\exp(-\alpha t^2 N/4)$, is a good approximation ($N/4$ is used as a scaling factor for $\alpha$).
The Figure shows no significant changes in the behaviour between $\lambda=0.9$ and $1.1$, even though these cases are at different sides of the critical point. There is instead a clear distinction from the plots for $\lambda=0$, for which universality is reached much faster (as had been noted in Section \ref{sec:gaussianregime}).

\bigskip

\medskip

\begin{figure}[!hbt]
\begin{center}
\includegraphics[width=0.45\textwidth]{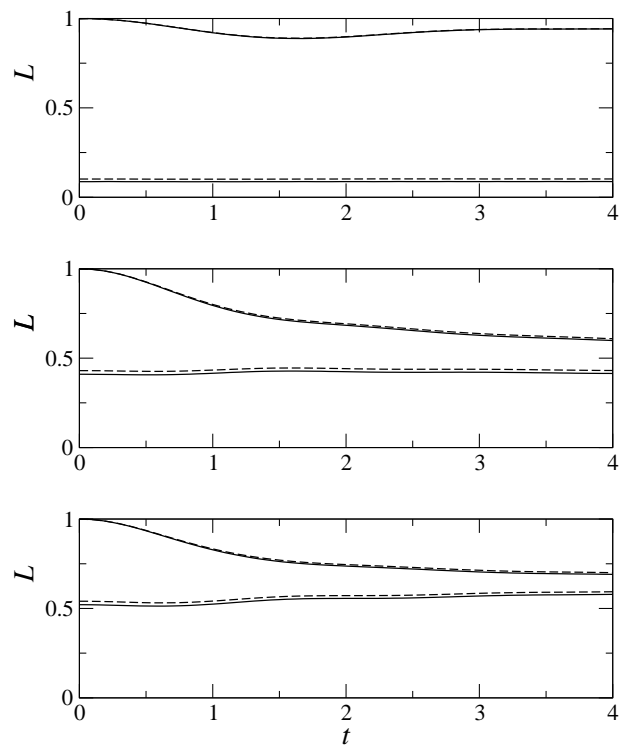}
\caption{The envelope of the echo as a function of time for a one-qubit system interacting  with one site of a spin chain with $N=100$, $\gamma=1$. The plot in the top shows the case \mbox{$\lambda=0.5$}, the middle corresponds to $\lambda=0.99$, and the bottom to \mbox{$\lambda=1.05$}. In each figure a full curve is plotted for \mbox{$g=50$}, and a dashed curve for $g=30$; both are very similar, regardless of the proximity to the critical point $\lambda=1$.}
\label{fig:env_univ}
\end{center}
\end{figure}

We can present other examples of similar situations, for instance, the case in which the one-qubit system is not central, but is inhomogeneously coupled to the different sites of the chain (a problem which has been treated in \cite{rossini-2007-75}): Figure \ref{fig:env_univ} shows the upper and lower envelopes of the echo for a qubit that interacts with only one site of the chain. In this case the environment has a phase transition but the condition for criticality is $\prod_i \lambda_i =1$ (for $\gamma=1$) \cite{Pfeuty79}. Then, even for large values of $g$ the interaction with the system may not be strong enough to drive the phase transition in the environment (in the homogeneous case, the situation was quite the opposite). Of the three cases in Figure \ref{fig:env_univ} only in the second one (\mbox{$\lambda=0.99$}) the perturbation can drive a phase transition. However, the fact that the curves are quite independent of $g$ is observed not only in that case but also in the other two ($\lambda=0.5$ --top-- and $\lambda=1.05$ --bottom). 

\vspace{45pt}

These results suggest that the existence and features of the universal regime are not clearly related to the phase transition. As this regime is reached for large coupling, it is natural to think that in this case, universality might instead be a consequence of a separation of time scales. In the strong-coupling limit, the fast oscillation (with frequency of order $g$) is thus given by the system-environment interaction, and the envelope is associated to the ``perturbation'' introduced by the chain Hamiltonian. In what follows we shall give a simple analytic derivation supporting this idea. For definiteness, we explain it in the central qubit case. We consider the evolution of the chain under the Hamiltonian $H_1=H_C-g Z_T$, with $H_C$ of order 1, $g\gg1$ and $Z_T=\sum_j Z_j$. We then decompose the state of the chain in the form:
\beq
\ket{\phi(t)}=e^{-iH_1t}\ket{E_0}=e^{igtZ_T}\ket{\phi'(t)}
\eeq
where $\ket{\phi'(t)}$ contains the slow evolution according to the equation:
\beq
\frac{d~}{dt}\ket{\phi'}=-i ~e^{-i gtZ_T} ~H_C~ e^{i gtZ_T} \ket{\phi'}
\eeq
which determines the time-dependent Hamiltonian in this interaction-like picture. The formal solution is thus given by:
\beq
\ket{\phi'(t)}= \exp\left(-i\int_0^t dt'~e^{-i gt'Z_T} H_C ~e^{i gt'Z_T}\right) \ket{E_0}.
\eeq
The exponent in the evolution operator for $\ket{\phi'}$ can be expanded in the computational basis of the chain, formed by the eigenstates of $Z_T$. These are noted as \mbox{$\ket{\vec k}=\ket{k_1 \ldots k_N}$}, with $k_j=0,1$, and have eigenvalues \mbox{$\Lambda_k=\sum_j (-1)^{k_j}$}. For $g\gg1$ we can approximate the exponent in the form:
\beqa
&\int_0^t ~\sum_{\vec k, \vec k'} ~e^{-i gt'(\Lambda_k-\Lambda_{k'})} ~(H_C)_{\vec k, \vec k'} ~\ket{\vec k}\bra{\vec k'} \simeq&\nonumber\\
&\simeq t ~\sum_{\vec k, \vec k'}~\delta_{\Lambda_k,\Lambda_{k'}} ~(H_C)_{\vec k, \vec k'} ~\ket{\vec k}\bra{\vec k'} ~= ~t~H'_C&
\eeqa
which means the slow evolution is, in the strong coupling limit, given by the chain Hamiltonian reduced to a block diagonal form $H'_C$, where blocks correspond to the different eigenvalues of the interaction term. This derivation can be repeated in more general situations, provided that all the energy differences $|\Lambda_k-\Lambda_{k'}|$ are either zero or over a finite gap. 
This approximation can now be inserted in the echo:
\beq
L(t) \simeq |\bra{E_0}e^{igtZ_T}e^{-itH'_C}\ket{E_0}|^2
\eeq
where we see that the evolution operator can be factorized into a fast oscillation with frequency of order $g$, and a slow evolution governed by an effective chain Hamiltonian $H'_C$. This slow evolution determines the envelope, and has a negligible dependence on $g$ (for large enough $g$). In this way, we have found an example of a universal regime that has no connection with phase transitions, and which is only due to energy-scale separation.

The approximation we have derived allows for short-time expansions of the envelope only. This can be useful in cases with no translation invariance, where the expressions of the echo are not as simple as (\ref{eq:mainformula}). For example, the case of one qubit interacting with one site of the chain shown in Figure \ref{fig:env_univ}. In the simplest approach, valid at very short times (compared to the chain evolution scales) we can consider the chain Hamiltonian only for the determination of the initial state $\ket{E_0}$, and neglect its effect in the subsequent evolution. The echo then takes the form:
\beq
L(t) \simeq 1- (1-\langle Z \rangle^2) \sin^2 (gt) \label{eq:shorttime1site}
\eeq
where the mean value is taken over one site, in the ground state of the isolated chain. The short-time behaviour is then determined by the magnetization in $z$ direction, which is somehow natural because this is the operator appearing in the perturbation. From this formula we learn that at short times the echo oscillates with frequency $2g$ taking values between 1 and $\langle Z \rangle^2$. We can thus see that the differences in the initial amplitudes of the envelopes in Figure \ref{fig:env_univ} are due to the increasing alignment of the spins with the external field. Further expansions of the operator $\exp(-itH'_C)$ allow in the same way for the description of the short-time decay of the envelope.

\section{Conclusions}
\label{sec:conclusions}

We have studied the decoherence induced on a one-qubit system by the interaction with a spin chain with XY Hamiltonian. For this purpose we have analyzed the Loschmidt echo, related to the decay of the off-diagonal terms in the reduced density matrix of the system. We have confirmed the existence of a regime of universal decoherence, namely, the fact that for large enough couplings the envelope of the echo becomes independent of the coupling intensity. However, we found that this behaviour does not provide a clear indication to assure the existence of a quantum phase transition in the environment. As conjectured in \cite{cucchietti-2006} a quantum phase transition may produce, quite generally, a universal 
regime of decoherence. But such a regime also appears in other situations. Even though the threshold for the interaction to reach the universal regime can vary depending on the problem under study, it is not clear either whether this threshold can be related to the phase transition. Furthermore, by simple analytic arguments we have shown that universality can be found whenever there is a large scale separation between the energies of the environment and the energies, and energy differences, of the interaction Hamiltonian. 

We have also analyzed some features of the time envelope of the echo for different values of the Hamiltonian parameters, corresponding to Gaussian and non-Gaussian decays. For the Gaussian envelope (found in the Ising case, with $\gamma=1, \lambda\simeq0$, and also in the short-time behaviour in more general cases), we have derived a new formula for the time width of the decay which is an improvement of the one presented in \cite{cucchietti-2006}. This new formula is in better agreement with the exact results for a wider range of values of $\lambda$: specially, it indicates that there is no enhancement of decoherence by approaching the quantum phase transition in the strong coupling regime. This is to be contrasted to the weak coupling case, where decoherence sharply increases in the proximity of the critical point. Throughout the article we have found more examples of this situation: we do not see a clear enhancement of decoherence at $\lambda\simeq1$ for a qubit interacting with only one site of the chain, nor for a central qubit when we take $\gamma$ close to 0. This suggests that the relation between decoherence and phase transitions is not the same in different coupling regimes.

\section{Acknowledgments}

We thank F. Cucchietti for fruitful discussions.

\end{document}